\documentstyle[aps,prl,epsfig]{revtex}

\newcommand{\be}{\begin{eqnarray}}
\newcommand{\ee}{\end{eqnarray}}

\begin{document}

\title{Cooper-pair coherence in a superfluid Fermi-gas of atoms}

\author{Gh.-S.~ Paraoanu$^{1}$\footnote{On leave from 
the Department of Theoretical Physics, National Institute
for Physics and Nuclear Engineering, PO BOX MG-6, Bucharest,
Romania.}, M. Rodriguez$^{2}$ and 
P.~T\"orm\"a$^{2}$}
\address{$^1$ Department of Physics,
University of Illinois at Urbana-Champaign, 1110 W. Green St., 
Urbana 
IL61801, USA \\
$^2$ Laboratory of Computational Engineering, P.O.Box 9400, 
FIN-02015
Helsinki University of Technology, Finland }

\maketitle

\date{\today}

\begin{abstract} 
  
  We study the coherence properties of a trapped two-component
  gas of fermionic atoms below the BCS critical temperature. We propose an
  optical method to investigate the Cooper-pair coherence across
  different regions of the superfluid. Near-resonant laser light is
  used to induce transitions between the two coupled hyperfine states.
  The beam is split so that it probes two spatially separate regions
  of the gas. Absorption of the light in this interferometric scheme
  depends on the Cooper-pair coherence between the two regions.

\end{abstract}


\section{Introduction}

The coherence properties of clouds of trapped bosonic atoms 
have been 
already the 
focus of several interesting experiments. It has been convincingly 
demonstrated that the single-component Bose-Einstein condensate
is coherent on the time-scale of most experiments 
\cite{singleBEC}, while
for two-state fluids the intra- and inter- component  coherence 
survives 
on remarkably long time-scales despite external manipulations 
which 
displace spatially the centers of mass of the two fluids with 
respect to 
each other \cite{doubleBEC}. 

In the case of trapped two-state Fermi gases with negative 
scattering 
length, the existence of a BCS transition has 
been predicted \cite{prediction}. By now, intensive experimental 
work on 
trapping and cooling fermionic atoms \cite{cold} with lower and 
lower 
fractions of Fermi temperature being reported and the success in 
trapping mixtures of bosons and fermions as well as two states 
of the same 
fermionic atom allow most of the 
researchers to be optimistic that the BCS critical temperature 
will soon 
be reached.

The main candidates for atoms to undergo the BCS transition are
$^{40}K$ and $^{6}Li$ \cite{cold}. For $^{6}Li$,
the states $|5\rangle$ and 
$|6\rangle$ as defined in \cite{houbiers}
have an anomalously large negative s-wave 
scattering length -- thus a relatively high critical temperature.
They have
also a small enough decay rate in the region of mechanical 
stability,
which gives the system a long enough lifetime to be 
experimentally 
interesting \cite{houbiers}. It is also assumed that the densities 
of the two 
components are close enough so that below the critical 
temperature one 
has the usual s-wave BCS coupling between the two states. 
We will denote the two states involved in pairing by 
$|\downarrow \rangle$ and $|\uparrow \rangle$.
The BCS 
transition is connected with the appearance of an order parameter 
\cite{degennes} 
$\Delta (\vec{r})= - \frac{4\pi\hbar^{2}a}{m} 
\langle\hat{\psi}_{\uparrow}(\vec{r})
\hat{\psi}_{\downarrow}(\vec{r})\rangle$. The ultraviolet divergences 
associated 
with using a contact interaction pseudopotential can be dealt, 
in case of metallic superconductors by 
introducing a cut-off at the Debye energy, and in case of
trapped atoms by other renormalization schemes, 
such as 
pseudopotentials regularized with the operator $\partial_{r}[r 
\times ]$ 
\cite{bruun}. The properties of the order parameter of cold 
fermionic gases 
in the superfluid phase have been studied in the 
regime close to the critical temperature, where it satisfies a 
Ginzburg-Landau
equation \cite{baranov}. 

In this paper we study the spatial coherence and pairing properties of 
the BCS-paired two-state Fermi gas at low temperatures, 
when the GL treatment is not applicable but instead one needs 
to use the Bogoliubov-deGennes formalism \cite{degennes}.
We propose a method for checking experimentally the coherence of 
Cooper pairs and measuring the Cooper pair size. The method 
is based on breaking some of the Cooper pairs by 
driving transitions 
between the states $|\uparrow \rangle$
and $|\downarrow \rangle$ with nearly-resonant light. 
The laser beam is split and the 
two resulting beams are focused on two spatially separated regions
of the condensate, after which they are merged together  
like in a typical interferometry experiment. If both the 
laser and the condensate coherence are preserved between the
two regions, interference should have an effect on the absorption of the
light. We show that this is indeed the case and the existence of 
interference contributions for a given beam separation shows that 
the gas has coherence on that length scale.  

Interaction of the superfluid gas with laser light has been considered
earlier as a method to probe the BCS transtition of the gas \cite{light,paivi}.
In these proposals the paired states were coupled to some unpaired
states of the atoms, and in the proposals \cite{light} the light was mainly
thought of being off-resonant. Here we consider laser-induced transitions
directly between the two states that are paired. The light is nearly resonant
and supposed to be absorbed in the pair breaking process. 
Moreover, the interferometric configuration
in our proposal allows to probe coherences, not only the
existence of the superconducting gap. 

We first describe the proposed method in section 2, then derive the
absorption rate for a general laser geometry in section 3. Absorption 
rate and the effect of coherences in the interferometric setup
are considered in section 4. The appendix discusses coherence
length scales in the BSC Green's functions.
 
\section{The idea}

The basic idea of coupling the two paired states by a laser
is depicted in Fig.1. The two states $|\uparrow \rangle$
and $|\downarrow \rangle$ are typically hyperfine ground states
thus one would actually use a Raman transition to couple them.
In the following we will, however, consider just laser light with
one Rabi frequency and detuning -- in case of a Raman transition 
these can be understood as effective quantities. Likewise,
Fig.1 also shows only one laser driving the transition.

One may make an educated guess that the laser has to have 
a finite detuning $\delta$ to drive the
transition. This is because one has to break the Cooper pair,
i.e.\ provide extra energy at least of the amount of twice the gap energy
in order to transfer and atom from, say, state $|\downarrow \rangle$ 
to state $|\uparrow \rangle$ which makes
both of the pairing partners to have the same internal state and
thus become excitations in the superfluid.
Our calculations show that this extra amount of energy is 
indeed required. 

The scheme for probing spatial coherence is shown in Fig.2.
The laser beam is split and focused into two regions with
a separation $z$. After this they have to be recombined,
or detected nonselectively -- this prevents from gaining
information about from which of the beams light was absorbed. 
This lack of information is essential for any interference phenomena 
to occur. Below we calculate, first for a general laser
configuration, the transfer rate of atoms
from one state to another -- which directly corresponds to 
the rate of absorption. Then we demonstrate the idea
by considering a specific interferometric configuration.

\section{The absorbtion rate -- general case}

In the rotating wave approximation the interaction of the laser light
with the matter fields can be described by a time-independent 
Hamiltonian in which the detuning $\delta$ plays the role of an
externally imposed difference in the chemical potential
of the two states. The total Hamiltonian becomes then $\hat{H} = 
\hat{\tilde{H}} 
+
\hat{H}_{T}$, where
\begin{equation}
\hat{\tilde{H}} = \hat{H}_{BCS} + \left(\mu + \frac{\delta}{2}\right)\int 
d\vec{r}
\hat{\psi}_{\uparrow}^{\dagger}(\vec{r})\hat{\psi}_{\uparrow}
(\vec{r})
+ \left(\mu - \frac{\delta}{2}\right)\int d\vec{r}
\hat{\psi}_{\downarrow}^{\dagger}(\vec{r})\hat{\psi}_{\downarrow}
(\vec{r}). 
\end{equation}
Here $\mu$ is the chemical potential of the Fermi gas before 
the laser was turned on, the same 
for both of the components in order to allow standard BCS pairing.
The Hamiltonian $\hat{H}_{BCS}$ is the BCS-approximation 
of the matter Hamiltonian with 
the chemical potential included \cite{degennes}. 
The transfer Hamiltonian is given by 
\begin{equation}
\hat{H}_{T} = \int d\vec{r}\Omega (\vec{r})\hat{\psi}_{\uparrow}
^{\dagger}(\vec{r})\hat{\psi}_{\downarrow}(\vec{r}) + \Omega^{*}
(\vec{r})\hat{\psi}_{\downarrow}^{\dagger}(\vec{r})
\hat{\psi}_{\uparrow}(\vec{r}),
\end{equation}
with $\Omega (\vec{r})$ characterizing the local strength of the 
matter-field interaction.  
We consider that at the time $-\infty$ the system was in its BCS ground state, 
$\Psi_{BCS}$. 
Then the laser field has been turned on. The intensity of the 
electromagnetic 
field is small enough for the transfer Hamiltonian $\hat{H}_T$ to 
be just a perturbation on $\hat{H}_{BCS}$. 

The main observable of interest is the 
rate of transferred atoms from, say, state $\downarrow$ to 
state $\uparrow$. This also directly corresponds to the
absorption of the light. It is defined 
\begin{equation}
I_{\uparrow} = \frac{\partial}{\partial t}\int d\vec{r}\langle\Psi 
(t)|\hat{\psi}_{\uparrow}^{\dagger}(\vec{r})\hat{\psi}_{\uparrow}(\vec{r})
|\Psi (t)\rangle 
\end{equation}
and can be further evaluated with the help of the Schr\"odinger equation
$i\hbar\frac{\partial}{\partial t}|\Psi (t)\rangle = \hat{H} |\Psi (t)
\rangle$ as
\begin{equation}
I_{\uparrow} = i \int d\vec{r}\langle\Psi (t)|
\Omega^{*}(\vec{r})\hat{\psi}_{\downarrow}^{\dagger}(\vec{r})
\hat{\psi}_{\uparrow}(\vec{r}) - \Omega (\vec{r})
\hat{\psi}_{\uparrow}^{\dagger}(\vec{r})
\hat{\psi}_{\downarrow}(\vec{r})|\Psi (t)\rangle.
\end{equation}
In the following we call $I_{\uparrow}$ the current in analogy to
metallic superconductors where the flux of electrons out of the
superconductor constitutes the electrical current.

We introduce an interaction representation with respect to 
$\hat{\tilde{H}}$ and use linear response theory with respect to $\hat{H}_{T}$  
\cite{mahan}. Validity of the linear response theory requires that
the laser intensity is small and the transfer of atoms can be
treated as a perturbation. This implies also that the amount of atoms
transferred from one state to another is not so big that it would 
severely imbalance the chemical potentials of the two species of atoms 
and thus break the BCS state.
In order to simplify notation we denote 
\begin{equation}
\hat{O}(t)
= e^{i\hat{H}_{BCS}t}\hat{O}e^{-i\hat{H}_{BCS}t},\label{repre} 
\end{equation}
where $\hat{O}$ is any operator.
The current becomes
\begin{equation}
I_{\uparrow} = \int d\vec{r}\int_{-\infty}^{t}
\langle\Psi_{BCS}|\left[\hat{j}(\vec{r},t), \hat{H}_{T}(t')\right]|\Psi_{BCS}
\rangle .
\end{equation}
Here for example
\begin{eqnarray}
\hat{H}_{T}(t) &=& \int d\vec{r}\Omega^{*}(\vec{r})e^{-i\delta t}
\hat{\psi}_{\downarrow}^{\dagger}(\vec{r},t)\hat{\psi}_{\uparrow}
(\vec{r},t) 
+ \Omega (\vec{r})e^{i\delta t}
\hat{\psi}_{\uparrow}^{\dagger}(\vec{r},t)
\hat{\psi}_{\downarrow}(\vec{r},t), \nonumber \\
\hat{j}(\vec{r},t) &=& \Omega^{*}(\vec{r})e^{-i\delta
t}\hat{\psi}_{\downarrow}^{\dagger}(\vec{r},t)
\hat{\psi}_{\uparrow}(\vec{r},t) - \Omega (\vec{r})e^{i\delta t}
\hat{\psi}_{\uparrow}^{\dagger}(\vec{r},t)
\hat{\psi}_{\downarrow}(\vec{r},t) \nonumber
\end{eqnarray}
are the transfer Hamiltonian and the current operator in the 
representation (\ref{repre}).
We introduce the operator $\hat{A} (\vec{r}, t) = 
\Omega^{*}(\vec{r})
\hat{\psi}_{\downarrow}^{\dagger}(\vec{r},t)\hat{\psi}_{\uparrow}(\vec{r},t)$;
using this the current of atoms to the $\uparrow$ state becomes
\begin{eqnarray}
I_{\uparrow} &=& \int_{-\infty}^{t}dt'\{ e^{i\delta 
(t'-t)}\int d\vec{r}\int 
d\vec{r}'\langle\Psi_{BCS}|\left[\hat{A}(\vec{r},t), 
\hat{A}^{\dagger}(\vec{r}',t')\right]|\Psi_{BCS}\rangle  \\ 
& & - e^{i\delta 
(t-t')}\int d\vec{r}\int
d\vec{r}'\langle\Psi_{BCS}|\left[\hat{A}^{\dagger}(\vec{r},t),
\hat{A}(\vec{r}',t')\right]|\Psi_{BCS}\rangle\}.
\end{eqnarray}
This expression can be calculated by introducing a Matsubara 
correlation function
\begin{equation}
X(i\omega) = - \int_{0}^{\beta}d\tau e^{i\omega\tau}\int d\vec{r}\int 
d\vec{r}'\langle{\cal T}_{\tau}\hat{A}(\vec{r},\tau 
)\hat{A}^{\dagger}(\vec{r}',0)\rangle
\end{equation}
which gives the rate of transfer by $I_{\uparrow} = -2 
Im\left[X_{ret}(-\delta ,\vec{r})\right]$ with the retarded correlation 
function defined by $X_{ret}(\omega ) \stackrel{i\omega\rightarrow 
\omega + i\epsilon}{=} X (i\omega )$. In the spirit of Wick's theorem, we 
split the Matsubara function into a part which contains only the 
superconductor Green's function ${\cal G}$  and a part which contains only 
the 
anomalous superconductor Green's funtion ${\cal F}$: $X = X_{{\cal F}} + 
X_{{\cal G}}$,
\begin{eqnarray}
X_{{\cal G}}(i\omega ) &=& \int_{0}^{\beta}d\tau e^{i\omega\tau}\int 
d\vec{r}\int 
d\vec{r}'\Omega^{*}(\vec{r})\Omega (\vec{r}'){\cal 
G}(\vec{r}',0;\vec{r},\tau 
){\cal G}(\vec{r},\tau; \vec{r}',0) \\
X_{{\cal F}} (i\omega ) &=& \int_{0}^{\beta}d\tau e^{i\omega\tau}\int
d\vec{r}\int 
d\vec{r}'\Omega^{*}(\vec{r})\Omega 
(\vec{r}'){\cal F}^{\dagger}(\vec{r},\tau; \vec{r}',0
){\cal F}(\vec{r},\tau; \vec{r}',0) .
\end{eqnarray}
After summation over Matsubara frequencies, and choosing
zero temperature and $\delta >0$, we get 
$I_{\uparrow} = 
I_{\uparrow ,{\cal G}} + I_{\uparrow ,{\cal F}}$ with 
\begin{eqnarray}
I_{\uparrow ,{\cal G}} &=& -2\pi\sum_{n,m}\left|\int d\vec{r}\Omega 
(\vec{r})v_{n}(\vec{r})u_{m}(\vec{r})\right|^{2}\delta (\epsilon_{n} + 
\epsilon_{m} - \delta ) \label{unu}\\
I_{\uparrow ,{\cal F}} &=& 2\pi\sum_{n,m}\int d\vec{r}d\vec{r}'\Omega
^{*}(\vec{r})\Omega  
(\vec{r}')u_{n}^{*}(\vec{r})u_{m}(\vec{r}')v_{m}^{*}(\vec{r})v_{n}(\vec{r}')\delta 
(\epsilon_{n} +
\epsilon_{m} - \delta ).\label{doi}
\end{eqnarray}
Here the triplet $(u_{n}, v_{n}); \epsilon_{n}$ is a solution of the 
(nonuniform) Bogoliubov-de Gennes equations \cite{degennes}
\begin{eqnarray}
\epsilon_{n}u_{n}(\vec{r}) &=& \left[\frac{\hbar^{2}}{2m}\vec{\nabla}^{2} 
+
V(\vec{r}) + U(\vec{r}) -\mu\right]u_{n}(\vec{r}) + \Delta 
(\vec{r})v_{n}(\vec{r})\\
-\epsilon_{n}v_{n}(\vec{r}) &=& \left[\frac{\hbar^{2}}{2m}\vec{\nabla}^{2} 
+ V(\vec{r}) + U(\vec{r}) -\mu\right]v_{n}(\vec{r}) - \Delta^{*}
(\vec{r})u_{n}(\vec{r}).
\end{eqnarray}
Here $V(\vec{r})$ is the external potential and $U(\vec{r}) = 
\frac{4\pi\hbar^{2}a}{m}\langle\hat{\psi}_{\uparrow}^{\dagger}(\vec{r})
\hat{\psi}_{\uparrow}(\vec{r})\rangle = \frac{4\pi\hbar^{2}a}{m}
\langle\hat{\psi}_{\downarrow}^{\dagger}(\vec{r})
\hat{\psi}_{\downarrow}(\vec{r})\rangle$ is the Hartree field.
Given the potential $V(\vec{r})$ one can solve the Bogoliubov-deGennes 
equations and insert the results into (\ref{unu})-(\ref{doi}). 

>From the equations (\ref{unu})-(\ref{doi}) on can immediately see
that given a non-trivial spatial dependence of the laser profile
$\Omega(\vec{r})$, the BCS coherence, that is, the spatial coherence of
$v_n(\vec{r})u_m(\vec{r})$ in different length scales 
will have an influence on the current.
In the following section we demonstrate this by a simple example.

\section{The absorption rate -- interferometric case}

We consider the setup of Fig.2, where the laser beam is split
and focused in the $z$-direction but homogeneous
in the two other directions, that is $\Omega (\vec{r}) 
= \Omega\left[\delta (z-z_{1}) + \delta (z-z_{2})\right]$.
This gives stronger $z$-dependence than focusing into two
points, and is probably also experimentally simpler.

In order to derive instructive results in closed form
we consider the case of a uniform gas.
This corresponds to working in a region of space 
much smaller than the oscillator length of the trap,
or having a shallow trap where confinement effects
are small. Solutions of the 
Bogoliubov-deGennes equations are given in this case by 
\begin{eqnarray}
u_{k} (\vec{r}) &=& u_k e^{ik\vec{r}} \quad ; \quad u_k^2 = \frac{1}{2}\left( 1 + 
\frac{\xi_{k}}{E_{k}}\right),\\
v_{k} (\vec{r}) &=& v_k e^{ik\vec{r}} \quad ; \quad v_k^2 = \frac{1}{2}\left( 1 - 
\frac{\xi_{k}}{E_{k}}\right),\\
E_{k} &=& \sqrt{\xi_{k}^{2} + \Delta^{2}}, 
~~~\xi_{k} = \frac{\hbar^{2}k^{2}}{2m}-\mu.
\end{eqnarray}

Cylindrical coordinates and momenta are used and the
sums are transformed into integrals.
The currents (\ref{unu})-(\ref{doi}) become 
\begin{eqnarray}
I_{\uparrow, {\cal G}} &=& -\frac{|\Omega |^{2}}{2}\int 
u^{2}_{k}v^{2}_{q}\left[
1 + \cos (k_{z} + q_{z})z\right]\delta (E_{q} + E_{k} - \delta ) \label{cur1} \\
I_{\uparrow, {\cal F}} &=& \frac{|\Omega |^{2}}{2}\int 
u_{k}v_{k}u_{q}v_{q}\left[
1 + \cos (k_{z} + q_{z})z\right]\delta (E_{q} + E_{k} - \delta ). \label{cur2}
\end{eqnarray}
Here $k = \sqrt{\rho^{2} + k_{z}^{2}}$, $q = \sqrt{\rho^{2} + 
q_{z}^{2}}$, $z=z_{1}-z_{2}$, and the integral symbol means $\int \equiv
\int_{0}^{\infty}d\rho
\rho
\int_{0}^{\infty}dk_{z}\int_{0}^{\infty}dq_{z}$. 
One can investigate how absence of coherence affects the results
by attaching to $u_{k} (\vec{r})$ and $v_{k} (\vec{r})$ phase factors
which describe random space- and time-dependent fluctuations and
by averaging over them. This makes the cosine-dependent term in $I_{\uparrow, {\cal G}}$
as well as the whole current $I_{\uparrow, {\cal F}}$ to disappear. 
Thus one can vary $z$ and see whether the current varies as well --- if the current
becomes constant for large enough $z$ one knows the oscillating terms are 
absent, that is, coherence 
is not preserved in that length scale. 

The currents $I_{\uparrow, {\cal G}}$ and $I_{\uparrow, {\cal F}}$ are presented
in Fig.3 and 4. In the numerics the $\delta$-functions in (\ref{cur1})-(\ref{cur2})
have been replaced by Lorentzians of width 0.01 to describe the finite laser linewidth.
It is interesting to compare the variation of the currents as a function
of $z$ to the typical length scale of Cooper pairing, the Cooper pair
size $x_{CP}$. The pair size can be estimated simply by the 
momentum--coordinate uncertainty relation, the uncertainty in momentum being 
determined by the corresponding uncertainty in energy (the gap).
This results to $x_{CP} = \frac{1}{\Delta}\sqrt{\frac{\mu}{2m}}$. 
In the case of Figs.3 and 4, $x_{CP} \sim 7$. One observes oscillations
occuring with approximately the period $\Delta z \sim 2$ which is of the
same order of magnitude as $x_{CP}$. For $z>x_{CP}$ the current gradually
becomes constant. 
The total current $I_{\uparrow, {\cal G}} + I_{\uparrow, {\cal F}}$
is shown in Fig.5. Exactly at $\delta=2\Delta$ the currents $I_{\uparrow, {\cal G}}$ 
and $I_{\uparrow, {\cal F}}$ cancel each other (this would be different in
a non-homogeneous case) but for larger values of detuning the current is
non-zero. The oscillations are visible. 

Another length scale involved is
the Fermi length $x_{F}$ given by $x_{F} = 1/\sqrt{2m\mu}$.
The relationship to the Cooper pair size is 
typically $x_{CP}\gg x_{F}$, in case of Figs.3 and 4 $x_{F}\sim 0.7 =0.1 x_{CP}$.
In physical units, $x_{CP} \sim 0.3 \mu m$ and $x_F \sim 0.03 \mu m$ in a system
with particle number $N \sim 10^7$ and trap frequency $\Omega = 2\pi \times 150 Hz$ 
(and the gap $\Delta = 0.1 E_F$). If $N \sim 10^5$, the
length scales are $x_{CP} \sim 2 \mu m$ and $x_F \sim 0.2 \mu m$. Thus for
realistic experiments, the Fermi length will always be below the diffraction
limit of light and cannot be resolved. This is the case for $x_{CP}$ too for
very high particle numbers and/or trap frequencies. But for moderate densities -- which
is expected to be the situation in the first experiments -- the Cooper pair size
may be of the order of several micrometers which can be resolved by light in principle.

Note that the detuning $\delta$ has bigger than twice the
gap to produce any current at all (there is some current below $2\Delta$ because
the $\delta$-functions in (\ref{cur1})-(\ref{cur2}) were replaced by Lorenzians).
Thus having or not having BCS pairing 
(nonzero or zero gap) makes a significant difference in the result, which 
means that the method can be used also to detect the onset of the phase 
transition, c.f.\ \cite{light,paivi,anna,all}. For certain types of 
perturbations the Cooper-paired atomic Fermi gas has also below-gap
excitations. A density perturbation term of the form $U(\vec{r},t) 
[ \hat{\psi}_{\uparrow}^{\dagger}(\vec{r})\hat{\psi}_{\uparrow}
(\vec{r}) + \hat{\psi}_{\downarrow}^{\dagger}(\vec{r})\hat{\psi}_{\downarrow}(\vec{r})]$
leads to appearance of a Bogoliubov-Anderson phonon \cite{anna}.
In our case the laser detuning gives a similar term in the
Hamiltonian, $\delta/2 [ \hat{\psi}_{\uparrow}^{\dagger}(\vec{r})\hat{\psi}_{\uparrow}
(\vec{r}) - \hat{\psi}_{\downarrow}^{\dagger}(\vec{r})\hat{\psi}_{\downarrow}(\vec{r})]$
(assuming that the laser is turned on and confined to a spatial region, one
may think of $\delta$ having space and time dependence). The minus sign,
however, leads to cancellation of certain terms in the response calculation
and to the absence of the usual Bogoliubov-Anderson phonon. This will be
discussed in detail in another publication \cite{mirta2}.

\section{Conclusions}

For standard metallic superconductors the coherence properties are
usually investigated by creating interfaces with normal metals or with
insulators. Optical manipulation of cold alkali gases offers the
unique opportunity of creating such interfaces at any point in space
and of controlling at will the transfer Hamiltonian across the
interface. We described a
method for testing BCS coherence in a system of atomic fermions cooled
below the critical temperature by driving laser-induced transitions
between the two paired hyperfine states in different regions of the
space. We presented numerical and analytical evidence for the
feasibility of the procedure. We show that the photon absorbtion rate
changes when the distance between the transition regions is
of the order of magnitude of the BCS correlation length. 
Absence of coherence would make the rate independent of the distance.
The method can
serve as well for the detection of the onset of BCS transition,
because the absorption peak is shifted to the detuning of twice
the gap energy.

\section{Acknowledgements} We thank the Academy of Finland for 
support
(projects 42588, 48845, 47140 and 44897). Gh.-S. P. also acknowledges
the grant NSF DMR 99-86199 and wishes to thank Prof. G. Baym and Prof. A. 
J. Leggett for useful discussions.

\section{Appendix}

The currents $I_{\uparrow, {\cal G}}$ and $I_{\uparrow, 
{\cal F}}$ are convolutions of 
pairs of correlation functions, thus their direct connection with 
$x_{CP}$ and $x_{F}$ is not transparent. In this appendix we discuss the
dependence of the BCS correlation funtions on these length scales.
It gives insight in the physics of Cooper-pair coherence and to the more
complicated length scale dependence of $I_{\uparrow, {\cal G}}$ and $I_{\uparrow, 
{\cal F}}$.

The Green's funtions for the superconductor are defined as
\begin{eqnarray}
{\cal F} (\vec{r},\tau; \vec{r}',\tau ') &=& - \langle{\cal 
T}_{\tau}\left[
\hat{\psi}_{\uparrow}(\vec{r},\tau)\hat{\psi}_{\downarrow}(\vec{r},\tau 
')\right]\rangle \\ 
{\cal F}^{\dagger} (\vec{r},\tau; \vec{r}',\tau ') &=& - \langle{\cal 
T}_{\tau}\left[
\hat{\psi}_{\uparrow}^{\dagger}(\vec{r},\tau)\hat{\psi}_{\downarrow}^{\dagger}
(\vec{r},\tau ')\right]\rangle\\
{\cal G} (\vec{r},\tau; \vec{r}',\tau ') &=& - \langle{\cal 
T}_{\tau}\left[
\hat{\psi}_{\uparrow}(\vec{r},\tau)\hat{\psi}_{\uparrow}^{\dagger}(\vec{r},\tau
')\right]\rangle .
\end{eqnarray}
In the uniform case and at zero temperature we get for the zero-time 
correlation functions
\begin{eqnarray}
{\cal G}(\vec{r},0; \vec{r}',0) &=& 
\frac{m}{4\pi^{2}r}\int_{-\mu}^{\infty}
d\xi\sin\left[\sqrt{2m(\mu +\xi )}r\right]\left(1 - 
\frac{\xi}{\sqrt{\xi^{2}+\Delta^{2}}}\right),\\
{\cal F}(\vec{r},0; \vec{r}',0) &=&
\frac{m}{4\pi^{2}r}\int_{-\mu}^{\infty} 
d\xi\sin\left[\sqrt{2m(\mu +\xi 
)}r\right]\frac{\Delta}{\sqrt{\xi^{2}+\Delta^{2}}},
\end{eqnarray}
where $r=|\vec{r}-\vec{r}'|$. Information about any possible coherence
length scales like $x_{CP}$ or $x_{F}= 1/\sqrt{2m\mu}$ is hidden 
in these equations. Note that for a normal system (zero gap)
the Fermi correlation has quite an explicite dependence 
on $x_{F}$:
\begin{equation} 
{\cal G}(\vec{r},0;\vec{r}',0)\stackrel{\Delta 
=0}{=}\frac{1}{2\pi^{2}x_{F}}\left(\frac{x_{F}^{3}}{r^{3}}\sin\frac{r}{x_{F}}
- \frac{x_{F}^{2}}{r^{2}}\cos\frac{r}{x_{F}}\right).
\end{equation}
The anomalous correlation function ${\cal F}$ is non-vanishing only when 
the gap is finite, reflecting the instability of the Fermi sea 
and the appearance of fermionic pairing. For ${\cal F}$ one can actually
derive an approximate formula which shows the dependence on $x_{CP}$ or $x_{F}$:
For the usual situation 
$\Delta \ll \mu$, and one can expand $\sqrt{2m(\mu + \xi ) r} \approx 
\frac{r}{x_{F}} + \frac{1}{2}\frac{r}{x_{CP}}\frac{\xi}{\Delta}$, which gives
\begin{equation}
{\cal F}(\vec{r},0; \vec{r}',0) = 
\frac{m\Delta}{2\pi^{2}r}\sin\left(\frac{r}{x_{F}}\right)
K_{0}\left(\frac{r}{2x_{CP}}
\right).
\end{equation}
Here $K_{0}$ is the zero order modified Bessel function which for a real 
argument $a$ admits the integral representation \cite{g}
\[
K_{0}(a) = \int_{0}^{\infty}\frac{\cos (ax)dx}{\sqrt{1 + 
x^{2}}}.
\]
For example, using the series expansion of $K_{0}$ we find 
that at $r$ small compared with $x_{CP}$ the anomalous correlation 
function decreases 
with $r$ as  $\ln(r/x_{CP})/r$,
\begin{equation}
{\cal F}(\vec{r},0; \vec{r}',0) = 
- 
\frac{m\Delta\sin\frac{r}{x_{F}}}{2\pi^{2}r}\left(\ln\frac{r}{4x_{CP}}
 + 0.577215\right).
\end{equation}

In conclusion, the spatial behavior of the normal and anomalous Green's 
functions is 
governed by two parameters, $x_{CP}$ and $x_{F}$. However, since 
$x_{F}$ is typically much smaller than the width of the laser beams 
we expect a dependence only on $x_{CP}$ in the result of a 
measurement.

\pagebreak

\begin{figure}
\caption{Laser coupling between the two paired states. Atoms in
two different hyperfine states $|\uparrow\rangle$ and $|\downarrow\rangle$ 
have a negative s-wave scattering length and
form a Cooper pair associated with the appearance of the superconductor
order parameter (gap) $\Delta $. The laser couples the states 
$|\uparrow\rangle$ 
and $|\downarrow\rangle$, the frequency of the laser is detuned from the atomic
transition frequency by the amount $\delta$. In order the transfer of the
atom from one state to other to happen the Cooper-pair has to be broken and
the atoms become excitations in the superfluid. Note that in a real experiment
a Raman scheme would be used instead of a single laser beam.}
\end{figure}
\begin{figure}
\caption{Schematic picture of the interferometric setup for probing coherences. 
The initial laser beam
is split coherently and focused on two spatial regions of the gas with the
separation $z$. After the beams have passed the gas they are recombined
again and the amount of absorption is measured.}
\end{figure}
\begin{figure}
\caption{The current $I_{\uparrow, {\cal F}}$ as a function of the separation
$z$ and the detuning $\delta$. The current oscillates with $z$, with the period
of $\sim 2 \sim x_{CP}/3$.
The current is peaked around $\delta = 0.2 = 2\Delta$. Here
the Fermi energy $\mu = 1$ and the gap $\Delta = 0.1$. All the variables
are dimensionless. For instance for the particle number $N=10^5$ and 
the trap frequency $\Omega = 2\pi \times 150 Hz$ the scales would be $[z] \sim 0.3 \mu m$
and $[\delta] \sim 75 kHz$.}
\end{figure}
\begin{figure}
\caption{The current $-I_{{\uparrow}{\cal G}}$. Oscillations similar to those in
$I_{\uparrow, {\cal F}}$ are observed. The parameters are like in Fig.3.}
\end{figure}
\begin{figure}
\caption{The total current $I = | I_{{\uparrow}{\cal G}} + I_{\uparrow, {\cal F}}|$. 
Oscillations are observed as in Figs.3 and 4.}
\end{figure}

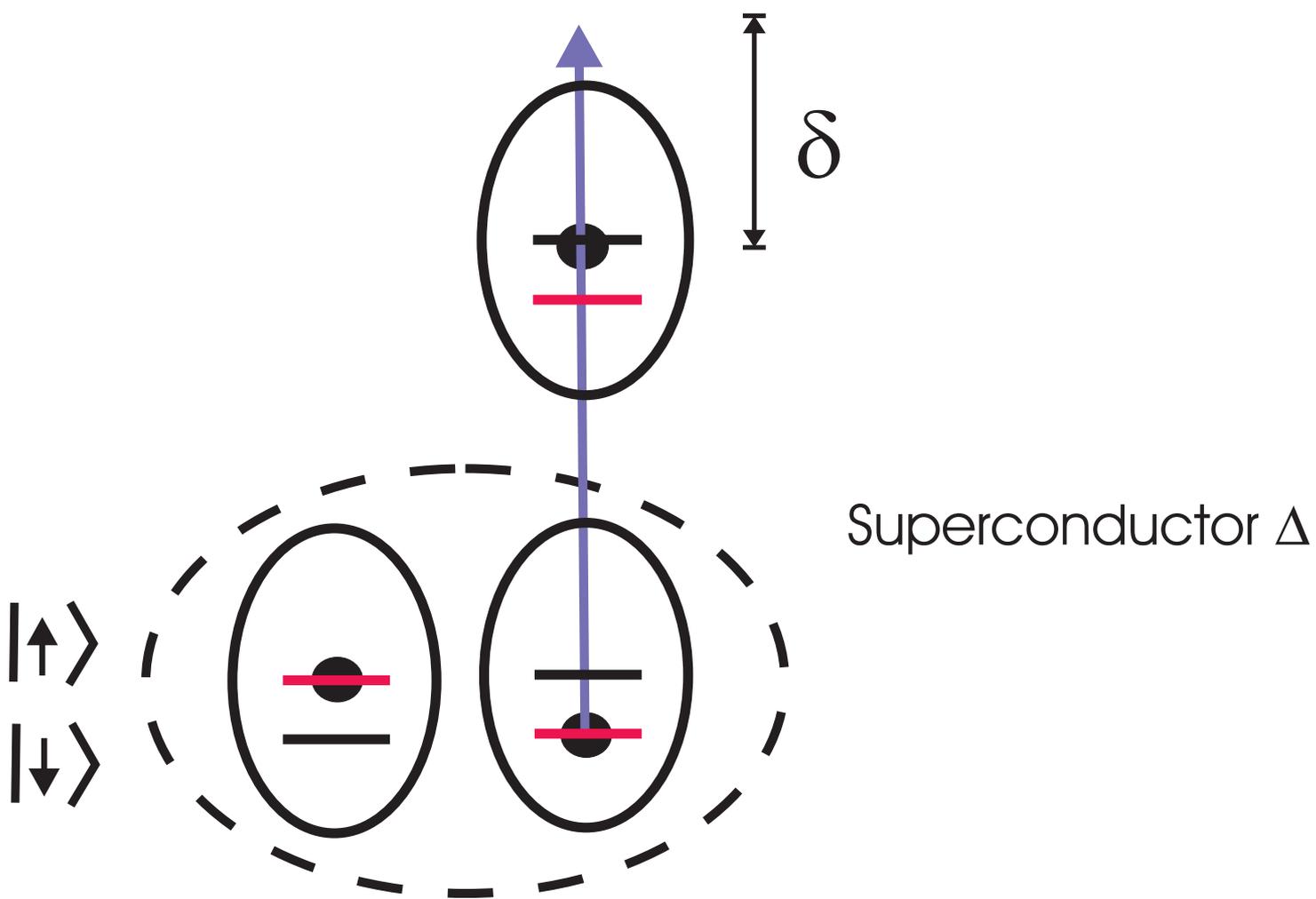
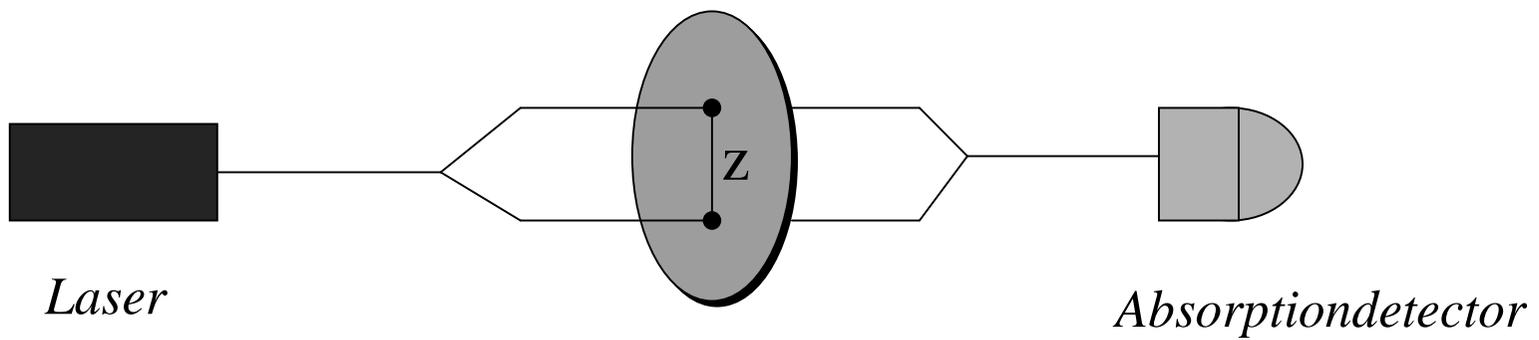
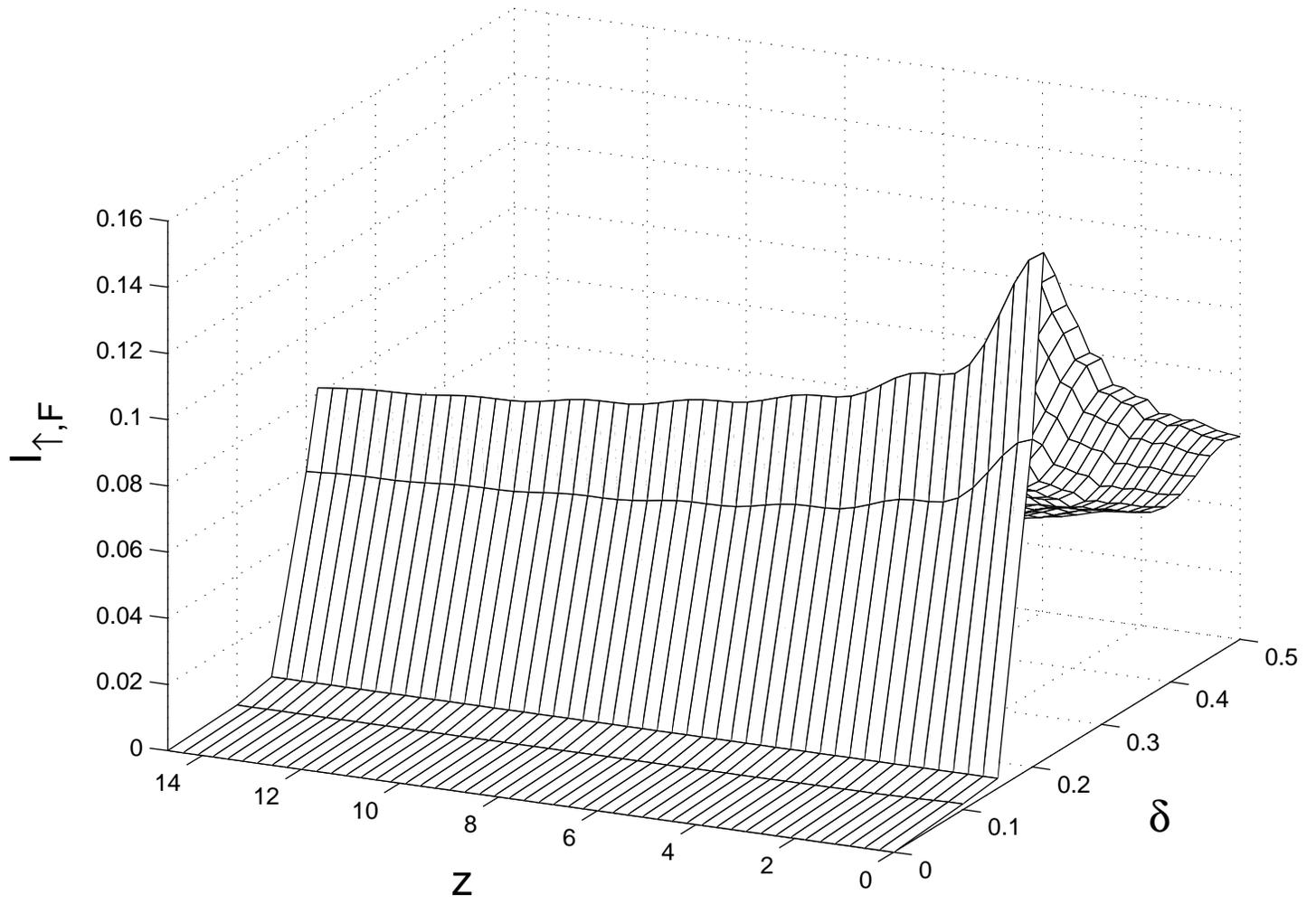
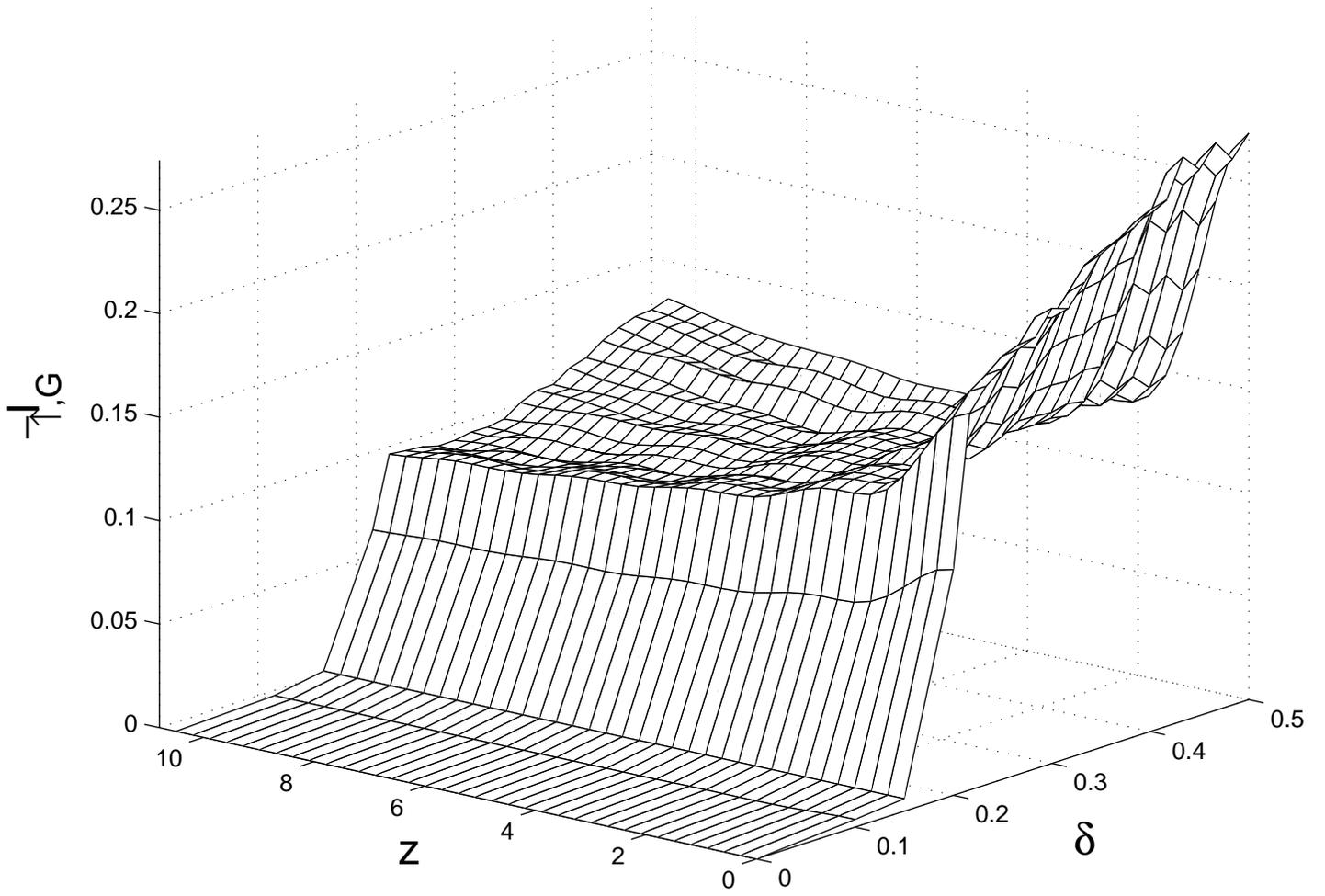
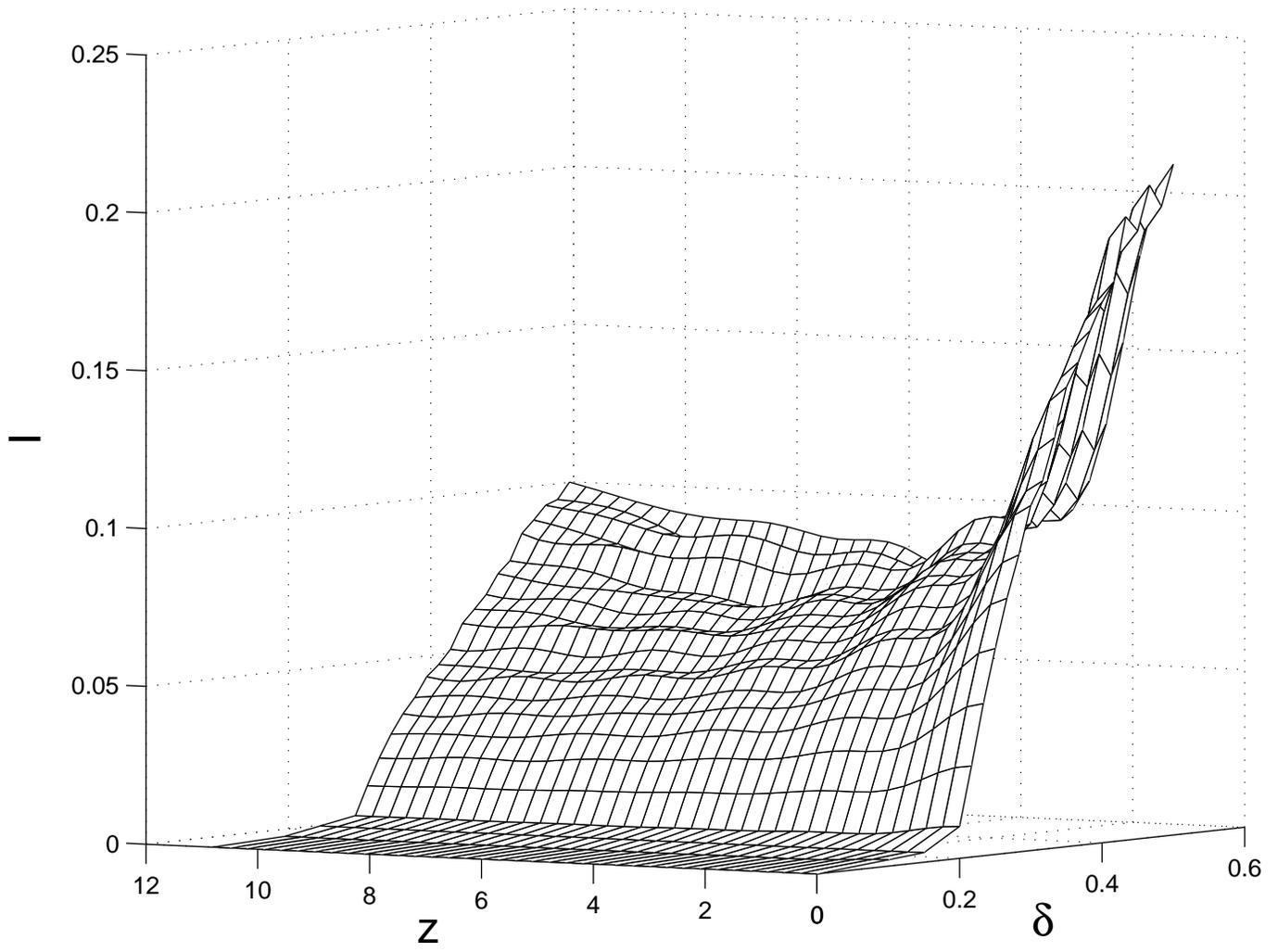
\end{document}